\documentclass[fleqn]{article}
\usepackage[margin=1.2in]{geometry} %
\usepackage{mathtools}
\usepackage{amssymb}
\usepackage{array}
\usepackage{booktabs}
\usepackage{newtxtext}
\usepackage{newtxmath}
\usepackage{mathrsfs}
\usepackage{dsfont}
\usepackage{cite}%
\usepackage{graphicx}
\usepackage{xcolor}
\usepackage{nicefrac}
\usepackage[hidelinks]{hyperref}
\usepackage[noabbrev]{cleveref}
\usepackage[nolist]{acronym}
\usepackage{xspace}
\usepackage[small]{caption}
\usepackage{subcaption}
\usepackage{mleftright}
\usepackage[symbol]{footmisc}
\usepackage{braket}
\mleftright
\makeatletter
\g@addto@macro\bfseries{\boldmath}
\newcommand*{\defeq}{\mathchoice{\mathrel{\rlap{%
                     \raisebox{0.24ex}{$\m@th\cdot$}}%
                     \raisebox{-0.24ex}{$\m@th\cdot$}}%
                     =}{\mathrel{\rlap{%
                     \raisebox{0.24ex}{$\m@th\cdot$}}%
                     \raisebox{-0.24ex}{$\m@th\cdot$}}%
                     =}{\mathrel{\rlap{%
                     \raisebox{0.08ex}{\small$\m@th\cdot$}}%
                     \raisebox{-0.28ex}{\small$\m@th\cdot$}}%
                     =}{\mathrel{\rlap{%
                     \raisebox{0.08ex}{\tiny$\m@th\cdot$}}%
                     \raisebox{-0.28ex}{\tiny$\m@th\cdot$}}%
                     =}}
\newcommand*{\eqdef}{\mathchoice{=\mathrel{\rlap{%
                     \raisebox{0.24ex}{$\m@th\cdot$}}%
                     \raisebox{-0.24ex}{$\m@th\cdot$}}}{%
					 =\mathrel{\rlap{%
                     \raisebox{0.24ex}{$\m@th\cdot$}}%
                     \raisebox{-0.24ex}{$\m@th\cdot$}}}{%
					 =\mathrel{\rlap{%
                     \raisebox{0.08ex}{\small$\m@th\cdot$}}%
                     \raisebox{-0.28ex}{\small$\m@th\cdot$}}}{%
					 =\mathrel{\rlap{%
                     \raisebox{0.08ex}{\tiny$\m@th\cdot$}}%
                     \raisebox{-0.28ex}{\tiny$\m@th\cdot$}}}%
                     }
\makeatother
\newcommand{\U}[1]{\ensuremath{\mathrm{U}(#1)}}
\newcommand{\dd}{\mathop{}\!\mathrm{d}}
\newcommand{\ii}{\hskip0.1ex\mathrm{i}\hskip0.1ex}

\newcommand{\WKS}{\ensuremath{S}}
\newcommand{\eff}{\mathrm{eff}}%
\newcommand{\tree}{\mathrm{tree}}%
\def\mytitle{Thermal History of Non-equilibrated Scalars}
\title{\mytitle}
\numberwithin{equation}{section}
\numberwithin{figure}{section}
\numberwithin{table}{section}
\begin{document}
\begin{titlepage}
\vspace*{1.0cm}

\begin{flushright}
UCI-HEP-TR-2025-09%
\\
\end{flushright}

\vspace*{2cm}

\begin{center}
{\Large\sffamily\bfseries\mytitle}

\vspace{1cm}

\renewcommand*{\thefootnote}{\fnsymbol{footnote}}

\textbf{%
V.\ Knapp--P\'erez\rlap{,}$^a$\footnote{vknappe@uci.edu} Gopolang Mohlabeng\rlap{,}$^{a,\,b,\,c}$\footnote{gmohlabe@sfu.ca} Michael Ratz$^a$\footnote{mratz@uci.edu} 
and Tim M.P.\ Tait$^a$\footnote{ttait@uci.edu}
}
\\[8mm]
\textit{$^a$\small
~Department of Physics and Astronomy, University of California, Irvine, CA 92697-4575 USA
}
\\[5mm]
  \textit{$^b$\small
~Department of Physics, Simon Fraser University, Burnaby, BC, V5A 1S6, Canada}
\\[5mm]
  \textit{$^c$\small
~TRIUMF, 4004, Westbrook Mall, Vancouver, BC, V6T 2A3, Canada}
\end{center}

\vspace*{1cm}

\begin{abstract}
Scalar fields in the early Universe are mostly discussed in two limits: either in equilibrium or completely decoupled.
In this work we discuss scenarios where there are scalar fields that are not in equilibrium, but for which the coupling to thermal bath leads to interesting non-trivial dynamics. 
For example, in theories where scalar fields control the effective couplings of the theory, such out-of-equilibrium behavior can lead to cases where the couplings vary during cosmological evolution. 
We systematically examine the generic features governing the evolution of these couplings, and as an application we highlight a novel effect where the scalar quartic coupling of an Abelian Higgs model is modified, leading to stronger cosmological phase transitions than would be obtained for static non-evolving quartics.
\end{abstract}
\vspace*{1cm}
\end{titlepage}
\renewcommand*{\thefootnote}{\arabic{footnote}}
\setcounter{footnote}{0}

\section{Introduction}

Weakly coupled scalar fields are a common ingredient in theories of physics beyond the Standard Model (SM).
An ubiquitous example is provided by string theories, where scalar fields referred to as moduli obtain \acp{VEV} which determine the effective coupling strengths of various interactions present in their effective field theory description.
These degrees of freedom are very weakly coupled, and thus are often expected to not reach thermal equilibrium in the early Universe --- in fact, their over-production is known to pose a threat to early cosmology~\cite{Coughlan:1983ci,deCarlos:1993wie,Kane:2015jia}. 
They are also known to respond to the dynamics of fields in the early Universe, e.g., by acquiring so-called Hubble masses~\cite{Dine:1995uk}. 
The fact that the couplings of the \ac{SM}  (and possibly other sectors)  are field-dependent gives rise to an effective thermal potential even for \emph{non-equilibrated} scalars. This has been demonstrated for gauge couplings~\cite{Buchmuller:2003is,Buchmuller:2004xr}, masses~\cite{Fardon:2003eh} and Yukawa couplings~\cite{Lillard:2018zts}. The impact of such varying effective couplings on the early Universe can be dramatic, with important consequences for, e.g.\ production of ultralight dark matter~\cite{Brzeminski:2020uhm,Batell:2021ofv,Heurtier:2021rko}, production of weakly-interacting massive particles \cite{Berger:2020maa,Howard:2021ohe}, baryogenesis \cite{Ipek:2018lhm,Ellis:2019flb,Croon:2019ugf}, leptogenesis~\cite{Croon:2022gwq,Bhalla-Ladd:2025agq}, and/or the Hubble tension~\cite{Kamionkowski:2024axz}. 

In this work, we systematically explore the cosmological dynamics of a scalar field (modulus) whose \ac{VEV} controls one or more couplings, in the limit where it is sufficiently weakly coupled to the thermal bath such that it remains out of thermal equilibrium, and yet its potential nonetheless acquires relevant thermal corrections. 
As a novel application, we study the consequences of a field-dependent scalar quartic coupling of a Higgs-like particle responsible for breaking an additional `dark' gauge symmetry we call $\U{1}^{\prime}$. 
We show that due to the thermal evolution of the modulus potential the relevant quartic coupling becomes smaller in the early Universe, thereby strengthening the resulting phase transition when the $\U{1}^{\prime}$ becomes spontaneously broken. 
Our  work is organized as follows: in \Cref{sec:WKS}, we summarize the general minimal thermal contributions from coupling scalars to various field configurations. In \Cref{sec:ModuliDependentPT}, we apply our analysis to a benchmark dark sector Higgs model. We show that the strength of the dark Higgs phase transition increases for a field-dependent quartic coupling. In \Cref{sec:Discussion}, we discuss the variation of the strength of the phase transition to initial conditions, outlining the validity of our work. We further discuss the consequences of this scenario to cosmological observables such as \ac{BBN} and primordial gravitational waves, as well as implications for string cosmology. Finally, we conclude in \Cref{sec:Conclusions}. The Appendix contains some technical details on the computation of the thermal effective potential for the dark Higgs model.

\section{Thermal Corrections to Weakly-coupled Scalar Fields}
\label{sec:WKS}

Our setup consists of a number of background particles with sufficiently strong interactions among them such that they
are in thermal equilibrium, described by a common temperature $T$. 
The thermal masses of the equilibrated plasma particles as a function of $T$ are computed in~\cite{Weldon:1982bn}. 
In addition, we include a number of scalar fields, which we generically denote as $S$, whose \acp{VEV} determine some or all of the strengths of the couplings among the plasma particles. The scalars S are sufficiently weakly coupled so as not to be in thermal equilibrium with the plasma.
Despite  themselves not making up an equilibrium population in the plasma, the effective potential of these weakly coupled scalars receives thermal contributions from the fact that the couplings and masses of the equilibrated plasma particles depend on their \ac{VEV}s. 

The free energy $\mathcal{F}$ is given at leading order by
\begin{align}
 \mathcal{F}&=\mathcal{F}(T,\WKS)\notag\\
 &\simeq-\frac{\pi^2}{30}g_*\,T^4+T^2\,\left(\sum_i c^{(g)}_i\, g_i^2(\WKS)
 +\sum_f c^{(y)}_f\,y_f^2(\WKS)+\sum_j c^{(m)}_j\,m_j^2(\WKS)
+\sum_{\ell}c_{\ell}^{(\lambda )}\, \lambda_{\ell} (\WKS)\right)\; , \label{eq:free_energy}    
\end{align} 
where the first term represents the free energy from the relativistic degrees of freedom, and the remaining terms are corrections from their finite masses and interactions.
Here, $g_i(\WKS)$ denote gauge couplings, $y_f(\WKS)$ Yukawa couplings, $m_j ( \WKS )$ masses and $\lambda_{\ell} ( \WKS )$ scalar quartic couplings, and  $g_*$ counts the (weighted) number of effectively massless equilibrated degrees of freedom in the plasma, 
\begin{equation}
  g_*=n_\text{bosons}+\frac{7}{8}n_\text{fermions}\;.
\end{equation} 
The dependence of the free energy on the couplings of the theory has been computed both in perturbation theory (beyond the leading orders) and on the lattice, see e.g.~\cite{Arnold:1994ps,Arnold:1994eb,Zhai:1995ac,Kajantie:2002wa}. 
For instance, for an $\text{SU}(N_c)$ gauge theory with $N_f$ Dirac fermions in the fundamental representation at one-loop,
\begin{equation}\label{eq:cg_in_SU(N)} 
  c^{(g)}=\frac{T^2}{64\pi^2}(N_c^2-1)\,(N_c+3N_f)\;.
\end{equation}
It turns out that for this case, the one-loop result approximates the dynamics obtained at higher order and the lattice rather well \cite{Kajantie:2002wa}.  
The fact that $c^{(g)}$ for this class of gauge interactions is positive-definite implies that an increase in the 
gauge coupling translates into an increase in the free energy.
This also turns out to be true for increases in Yukawa interactions~\cite{Lillard:2018zts} and the mass of the particles in a thermal bath~\cite{Fardon:2003eh},
\begin{subequations}
\begin{align}
  c_f^{(y)}&=\frac{5}{576}T^2\;,\label{eq:cy}\\
  c_j^{(m)}&=\frac{1}{24}\;,\label{eq:cm}\\
  c_{\ell}^{(\lambda )} &= \frac{3}{576}T^2\label{eq:clambda}\;.
\end{align}  
\end{subequations} 
Here, the coefficient $c_f^{(y)}$ is taken from~\cite[Appendix~C]{Lillard:2018zts},  $c_j^{(m)}$ can be obtained following~\cite{Fardon:2003eh} by expanding in $T$, and  $c_{\ell}^{(\lambda)}$ was computed in~\cite[Equation (3.78)]{Laine:2016hma} at the two-loop level. 
\Cref{eq:free_energy} accounts for all renormalizable couplings, with the exception of a scalar cubic term, which is absent in the \ac{SM} in the unbroken electroweak phase. 
The 2-loop contribution to the free energy from the cubic term has been computed in \cite{Laine:2017hdk}.
In our discussion, we focus on the renormalizable interactions $g$, $y$, $m$ and $\lambda$, noting that all of the coefficients in \eqref{eq:free_energy} are positive. As a rule of thumb the free energy ``wants'' couplings and masses to be as small as possible\rlap{,}\footnote{Similarly to the discussion of symmetry restoration at high temperature (cf.\ e.g.~\cite{Jansen:1998rj}) it is possible to construct counter-examples to this statement, cf.\ our discussion in \Cref{subsec:More_sophisticated_scenarios}.} 
favoring lighter, more weakly-interacting elements in the plasma.

The overall picture for the cosmology of moduli whose values control the sizes of renormalizable couplings and masses can be summarized as follows:
\begin{enumerate}
 \item The moduli are weakly coupled scalar fields, denoted collectively by $\WKS$, the \acp{VEV} of which determine the couplings and/or masses of some of the equilibrated particles. 
 We emphasize that $\WKS$ are \emph{not} in thermal equilibrium.
 \item The thermal bath exerts a force on $\WKS$, described by an $\WKS$-dependent thermal potential given by the free energy in \Cref{eq:free_energy}. 
 \item The effective thermal potential of $\WKS$, which is not to be confused with the thermal potential of the equilibrated particles, generically tends to drive $\WKS$ to values at which the couplings and masses \emph{decrease} at high temperature.
\end{enumerate}

We focus on the quartic coupling, whose field-dependence tends to enhance the strength of the associated phase transition. For the purposes of this work, we will leave the other renormalizable couplings $\WKS$-independent.

\section{Phase Transitions with Field-dependent Couplings}
\label{sec:ModuliDependentPT}

In this Section, we apply the picture in \Cref{sec:WKS} to the quartic coupling of a (dark sector) Higgs scalar whose \ac{VEV} spontaneously breaks a $\U{1}^{\prime}$ gauge symmetry.\footnote{Recently, it has been pointed out that if the quadratic term in the effective potential increases with decreasing temperature, this can result in a stronger phase transition~\cite{Chaudhuri:2025ybh}. The scenario we discuss here only relies on field-dependent couplings, and these effects decrease when the temperature becomes smaller.}
The Lagrange density is given by
\begin{equation}
    \mathscr{L}  = -\frac{1}{4}F^{\prime}_{\mu\nu} F^{\prime \mu \nu} + \left( D_\mu h\right)\,\left( D^{\mu}h\right)^{*} + \mu^2\, h^{*}h- \lambda(\WKS)\,\bigl( h^{*}h\bigr)^2  + \frac{1}{2} (\partial_\mu S) (\partial^\mu S) - \mathscr{V}_{\WKS} \; .
    \label{eq:LbeforeSB}
\end{equation}
Here the Higgs field $h$ is a (thermalized) complex scalar field with charge normalized to $q=1$ under the $\mathrm{U}(1)^{\prime}$ gauge symmetry and $F_{\mu\nu}^{\prime}$ is the field strength tensor of the $\mathrm{U}(1)^{\prime}$ (thermalized) gauge field $A_{\mu}^{\prime}$. 
In \eqref{eq:LbeforeSB}, the covariant derivative is $D_\mu = \partial_\mu - \ii  g\, A_{\mu}^{\prime}$ with the gauge coupling $g$. 
The Higgs quartic $\lambda(S)$ in \eqref{eq:LbeforeSB} depends on the modulus $S$ as
\begin{equation}
\lambda(\WKS) = \left(\frac{\WKS}{\Lambda}\right)^2  \;,
\label{eq:Quartic-Moduli}
\end{equation}
where $\Lambda$ represents the cut-off scale of some unspecified high energy physics responsible for generating the quartic, which we assume is large enough to ensure that $\WKS$ is not in thermal equilibrium with the elements of the plasma.
We assume that $\WKS$ is stabilized at some value $S_0$ by an external potential, with leading term given by 
\begin{equation}\label{eq:V_S}
 \mathscr{V}_{\WKS}=\frac{m_{\WKS}^2}{2}\bigl(\WKS-\WKS_0\bigr)^2\;.
\end{equation}
Generally, there will be higher-order terms in \Cref{eq:Quartic-Moduli,eq:V_S}, but they can be chosen to be small, and are unlikely to disrupt the qualitative picture resulting from approximating the potential as \Cref{eq:Quartic-Moduli,eq:V_S}, and thus we neglect them below.

At high temperatures, the \ac{VEV} $\braket{h}$ depends on the temperature $T$ and is the minimum of the thermal effective potential $\mathscr{V}_\eff^{(h)}(v, T,\mu, \lambda, g )$,
\begin{equation}
    \frac{\partial \mathscr{V}_\eff^{(h)}(v, T,\mu, \lambda, g )}{\partial v}\bigg|_{v=\braket{h}}=0 \; .
\end{equation}
We detail the effective potential for the equilibrated Higgs in \Cref{eq:VeffAppendix}. 
For small enough $\lambda$, the transition from a broken symmetry phase with $\braket{h}\neq 0 $ to a restored symmetry phase  with $\braket{h} = 0 $ is a first-order phase transition. Furthermore, we find that the phase transition becomes stronger for smaller $\lambda$, matching results obtained in the literature (see e.g.\ Figure~5 in~\cite{Jansen:1985nh}).

From \Cref{eq:2-loop}, we see that the free energy of the field $\WKS$ receives contributions from the Higgs quartic coupling $\lambda$. 
Since the quartic coupling depends on the field $\WKS$, this leads to a thermal contribution to the potential of $\WKS$. 
At leading order, this contribution is given by
\begin{equation}
    \mathscr{V}_{\eff,T}^{(\WKS)}(\WKS, T, \Lambda) = 2 \times \frac{3}{4}\left( \frac{\WKS}{\Lambda} \right)^2\, \frac{T^4}{144}\; .
    \label{eq:VeffPhiThermal}
\end{equation}
The factor of $2$ in the right-hand side of \Cref{eq:VeffPhiThermal} stems from the fact that a complex scalar has 2 degrees of freedom.
Using this information, we solve the equation of motion for $\WKS$ in an expanding Universe, which is given by
\begin{equation}
    \ddot{\WKS}(t) + 3H(t)\,\dot{\WKS}(t) + \frac{\partial \mathscr{V}^{(\WKS)}_{\eff}(\WKS, T,\Lambda,m_\WKS, \WKS_0)}{\partial\WKS} = 0 \; .
    \label{eq:EqMotionphi}
\end{equation}
Here, the dot represents the derivative with respect to time, with the full effective potential given by
\begin{equation}
    \mathscr{V}^{(\WKS)}_{\eff}(\WKS, T,\Lambda,m_\WKS, \WKS_0) = \frac{1}{2}m_\WKS^2\, \left(\WKS - \WKS_0 \right)^2+ \mathscr{V}_{\eff,T}^{(\WKS)}(\WKS, T, \Lambda)\; ,
    \label{eq:Veffphi}
\end{equation}
where $\WKS_0$ denotes the location of the zero-temperature minimum of $\WKS$. 
At high temperatures, the second term in \Cref{eq:Veffphi} dominates, driving $\WKS$ to small values for which the quartic coupling $\lambda = \WKS^2 / \Lambda^2$ is reduced. 
As the Universe evolves and the temperature decreases, the field is driven to its zero-temperature minimum $\WKS_0$, yielding a larger quartic coupling. 
As a result, in this scenario the quartic coupling is effectively shifted to a smaller value at the time (temperature) for which it crosses the first-order phase transition line, yielding a stronger phase transition compared to the scenario with a constant quartic coupling. . 

We numerically solve \eqref{eq:EqMotionphi} for some representative benchmark values to qualitatively study this behavior. 
It is instructive to write \Cref{eq:EqMotionphi} in terms of $T$, assuming $\dot{T} \approx -H\,T$ in a radiation-dominated Universe, yielding
\begin{equation}
    H^2(T)\, T^2\WKS^{\prime\prime}(T) + \frac{\partial \mathscr{V}^{(\WKS)}_{\eff}(\WKS, T,\Lambda,m_\WKS, \WKS_0)}{\partial\WKS} = 0\; ,
    \label{eq:PhiMotionT}
\end{equation}
where the prime denotes the derivative with respect to temperature. 
We impose initial conditions
\begin{subequations}\label{eq:InitialConditions}
\begin{align}
    \WKS(T_\mathrm{ini}) & = \WKS_{\mathrm{min}}(T_\mathrm{ini},\Lambda,m_\WKS,\WKS_0)\;, \\
    \WKS^{\prime}(T_\mathrm{ini}) &= 0\; ,
\end{align}
\end{subequations}
where $\WKS_{\mathrm{min}}(T_\mathrm{ini},\Lambda,m_{\WKS},\WKS_0)$ is the instantaneous minimum of $\mathscr{V}^{(\WKS)}_{\eff}(\WKS, T,\Lambda,m_\WKS, \WKS_0)$ at some initial temperature $T=T_\mathrm{ini}$.   
In other words, we assume that the field starts out at $T_\mathrm{ini}$ at the minimum of its effective potential at that temperature, and with no kinetic energy.  We explore perturbations to this assumption in
\Cref{subsec:Sensitivity} below, where we show that the results are not strongly dependent on these assumptions.

\begin{figure}[t]
    \centering
    \includegraphics{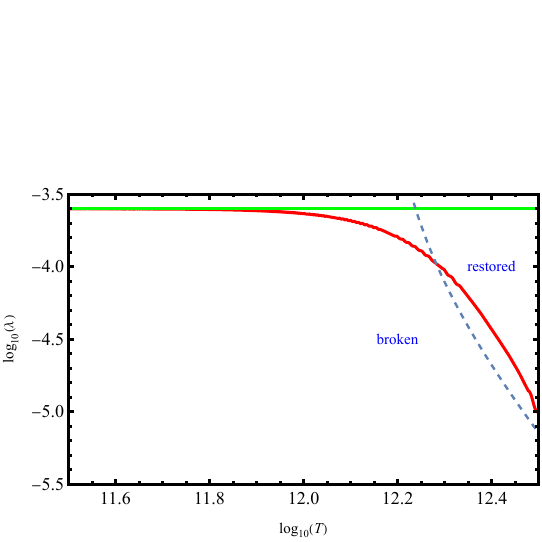}
    \caption{Phase diagram of an Abelian Higgs coupled to a $\mathrm{U}(1)$ gauge boson with benchmark values $g=0.3$, $\mu = 1.3\times 10^{11}\,$GeV, $\Lambda = 10^{15}\,$GeV, $m_{\WKS} = 7.2 \times 10^{8}\,$GeV, $\WKS_0 = 10^{-3.6/2} \times \Lambda$ and $T_\mathrm{ini} = 3.16 \times 10^{12}\,$GeV. 
    The blue dashed line demarks the first-order phase transition boundary.   The green line represents the trajectory of the Universe assuming a constant Higgs quartic coupling $\lambda = 10^{-3.6}$ coupling, i.e.\ a coupling that is not field-dependent. 
    By contrast, the red line represents the trajectory of the Universe if the quartic coupling is given by $\left( \WKS / \Lambda \right)^2$.
    While the late-time behavior of both cases coincides, we see that the dynamics at the time of the phase transition is quite sensitive to the field-dependence of $\lambda$.}
    \label{fig:PT-Moduli}
\end{figure}

In \Cref{fig:PT-Moduli}, we present the results of solving the equation of motion in \eqref{eq:PhiMotionT} to obtain the dynamics of $\lambda (\WKS)$. 
The blue dashed line represents the first-order phase transition boundary, and the green line represents the trajectory the Universe takes as it cools down for a constant (corresponding to $m_S \rightarrow \infty$) value of $\lambda = 10^{-3.6}$. 
It is worth noting that for large values of $\lambda$, the perturbative expansion of the effective potential in \Cref{eq:VeffAppendix} becomes unreliable, and thus the blue dashed line cannot be consistently extended into that region.
In red, we plot the trajectory of the Universe if the Higgs quartic coupling is given as in \Cref{eq:Quartic-Moduli} for the benchmark values $g=0.3$, $\mu = 1.3\times 10^{11}\,$GeV, $\Lambda = 10^{15}\,$GeV, $m_{\WKS} = 7.2 \times 10^{8}\,$GeV, and $T_\mathrm{ini} = 3.16 \times 10^{12}\,$GeV. 
As the Universe cools down, the solution approaches its zero-temperature value $\left( \WKS_0 / \Lambda \right)^2 = 10^{-3.6}$, but for higher temperatures
the influence of $S$, driven by the thermal corrections to its potential, results in a smaller effective quartic.

\begin{figure}[t]
  \centering\subcaptionbox{Contour plot.\label{fig:Scan}}
  {\includegraphics{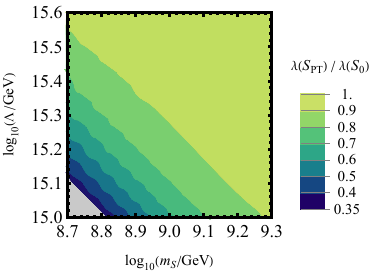}}\qquad
  \subcaptionbox{Fit.\label{fig:Fit}}
  {\includegraphics{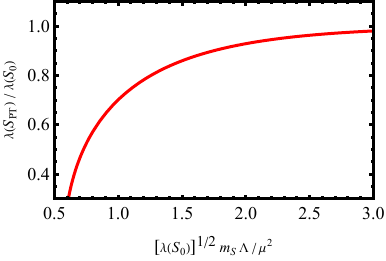}}
  \caption{(a) Contours of the ratio $\lambda\left(\WKS_{\mathrm{PT}}\right)/\lambda\left(\WKS_{0}\right)$ as a function of the mass $m_{\WKS}$ and the scale $\Lambda$. Here, $\lambda\left(\WKS_{\mathrm{PT}}\right)$ is the quartic coupling evaluated at the point where the field crosses the first-order phase transition line while $\lambda\left(\WKS_{0}\right)$ corresponds to its zero-temperature value. We fix $g=0.3$, $\mu = 1.3 \times 10^{11}\,$GeV and $\WKS_{0} = 10^{-3.6/2}\times \Lambda\,$, as in \Cref{fig:PT-Moduli}, yielding $\lambda\left(\WKS_{0}\right) = 10^{-3.6}$. The gray region represents the area where the quartic coupling $\lambda\left(S\right)$ starts in the broken phase. 
  (b) Numerical fit to the contours shown in panel (a), showing the ratio $\lambda\left(\WKS_{\mathrm{PT}}\right)/\lambda\left(\WKS_{0}\right)$ as a function of the dimensionless quantity $\left[\lambda\left( \WKS_0 \right) \right]^{1/2} m_S\, \Lambda / \mu^2$.}
  \label{fig:ScanAndFit}    
\end{figure}

The strength of the first-order phase transition depends sensitively on the mass of the moduli $m_{\WKS}$ and the cut-off scale $\Lambda$. 
We show this dependence in \Cref{fig:Scan}, where we present the contours of the ratio $\lambda\left(\WKS_{\mathrm{PT}}\right)/\lambda\left(\WKS_{0}\right)$ in the 
$m_{\WKS}$-$\Lambda$ plane. 
Here, $\WKS_{\mathrm{PT}}$ denotes the modulus field at the temperature where the quartic coupling crosses the first-order phase transition line. 
$\WKS_{0}$ corresponds to the value of the modulus at its zero-temperature minimum, cf.\ \Cref{eq:Veffphi}. 
We use the same parameters as in \Cref{fig:PT-Moduli} with the same initial conditions, as fixed by \Cref{eq:InitialConditions}. 
Note that the value of $\WKS_0$ changes accordingly with $\Lambda$ to maintain a zero-temperature value for the quartic coupling of $\lambda\left( \WKS_{0}\right) = 10^{-3.6}$. 

As $m_{\WKS}$ or $\Lambda$, or, more precisely, their product $m_{\WKS}\,\Lambda$, increases, the ratio becomes closer to one, indicating a weaker enhancement of the phase transition. 
This behavior is illustrated in \Cref{fig:Fit}, which presents a numerical fit to the contours displayed in \Cref{fig:Scan}. 
The ratio $\lambda\left(\WKS_{\mathrm{PT}}\right)/\lambda\left(\WKS_{0}\right)$ is approximately given by a function of the dimensionless quantity $\left[\lambda\left( \WKS_0 \right) \right]^{1/2} m_S\, \Lambda / \mu^2$. 
For values exceeding $2$, the moduli-dependent quartic coupling starts to become indistinguishable from its zero-temperature counterpart. 
In contrast, values below $2$ yield a pronounced enhancement, indicating a stronger first-order phase transition. 
The fit employs the ansatz
\begin{equation}
\label{eq:ansatz}
\frac{\lambda \left( \WKS_{\mathrm{PT}}\right)}{\lambda \left( \WKS_0 \right)} = \left( 1 - a\,\exp\left[-b\,\left[\lambda\left( \WKS_0 \right) \right]^{1/2} m_S \Lambda / \mu^2 \right]\right)^{\alpha}\;,
\end{equation}
with best-fit parameters $a = 2.0$, $b = 1.3$  and $\alpha = 0.41$.
This behavior is consistent with the expectation that increasing either $m_{\WKS}$ or $\Lambda$ decreases the thermal contribution relative to the zero-temperature term of the potential for $\WKS$ in \Cref{eq:Veffphi}. 
As a result, the value of $\lambda \left( \WKS_{\mathrm{PT}}\right)$ approaches its zero-temperature value. 

For small enough values of $m_{\WKS}$ and $\Lambda$, the initial value of the quartic coupling $\lambda\left(S_{\mathrm{ini}}\right)$ causes the Universe to begin in a broken phase.
This area is shown in gray in \Cref{fig:Scan}. 
However, in \Cref{sec:EffectivePotentialHiggs}, we assume that the dark Higgs and the dark photon are sufficiently light to stay in thermal equilibrium, which may no longer hold in the broken phase, and this region of the plot does not correspond to a meaningful result for $\lambda\left(\WKS_{\mathrm{PT}}\right)/\lambda\left(\WKS_{0}\right)$.
Over-all, we find that the strength of the quartic coupling at the phase transition can decrease by more than a factor 2, for appropriate choices of the moduli mass and coupling $\Lambda$.

Altogether, we thus see that the Universe undergoes a stronger first-order transition if the quartic coupling is field-dependent, which may have important implications for phenomena that rely on out-of-equilibrium dynamics such as baryogenesis, and observational effects such as production of a stochastic background of gravitational waves.

\section{Discussion}
\label{sec:Discussion}

We have shown that a field-dependent quartic leads to a stronger first-order phase transition. In what follows, we briefly discuss some of the possible constraints, observables, and implications of our findings. 

\subsection{Constraints and Observational Effects}
\label{subsec:Constraints}

Weakly coupled scalars can be a threat to successfully realizing the cosmology in the early Universe  \cite{Coughlan:1983ci,deCarlos:1993wie,Kane:2015jia}. 
For instance, their late-time decays may spoil the successful predictions from \ac{BBN}, dilute a baryon asymmetry of the Universe to unrealistic values, or distort the measured properties of the cosmic microwave background.  In fact, in some cases these constraints become much more severe as a result of the presence of the thermal effects we discuss here (see e.g.~\cite{Lillard:2018zts} for an example of the constraints from \ac{BBN} when the Yukawa couplings are field-dependent). 
Moreover, scenarios of low-temperature baryogenesis are also subject to constraints from collider experiments~\cite{Bittar:2024fau,Bittar:2024nrn}. In such scenarios, the relevant couplings may be field-dependent, potentially resulting in a link between the physics of the modulus and collider observations.

Our numerical analysis of the basic module in \Cref{sec:ModuliDependentPT} has focused on rather large mass scales $\gtrsim 10^8$~GeV. 
This choice is driven by a number of considerations; one reason is to avoid constraints from \ac{BBN} from late-time decays of $\WKS$.  
The decay width of $\WKS$ scales as $\Gamma_\WKS \sim m^3_S\, S^2_0 / \Lambda^4$, which must result in $\WKS$ decaying far before the time of \ac{BBN} characterized by $T_{\mathrm{BBN}} \sim$~MeV. 
Another motivation to consider large mass scales is because they can provide a direct link to visible sectors, thus providing further possible observational evidence for a strong first order phase transition. For instance, $\U1^{\prime}$ could be related to $\U1_{B-L}$, which can naturally explain the observed small active neutrino masses via the seesaw mechanism, if broken at a sufficiently high mass scale~\cite{Mohapatra:2005wg}. At the same time, breaking this symmetry, in the presence of a modulus may lead to a strongly first order phase transition, generating primordial gravitational waves which are detectable by near-future experiments, see~\cite{Hasegawa:2019amx} for further details. 

That said, it is worth mentioning that nothing particularly prevents migrating our discussion to smaller scales. 
For instance, $\WKS$ may decay to a dark sector to which it has larger couplings, leading to possible interesting consequences for well motivated sectors such as dark photons and/or dark matter, among many others. We leave quantitative exploration of these possibilities for future dedicated studies.

\subsection{Sensitivity to Initial Conditions}
\label{subsec:Sensitivity}

\begin{figure}[t]
    \centering
    \includegraphics{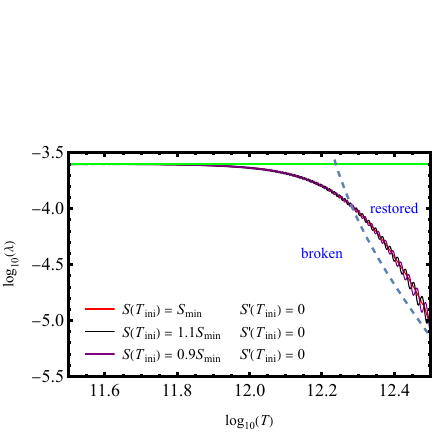}
    \hfill
    \includegraphics{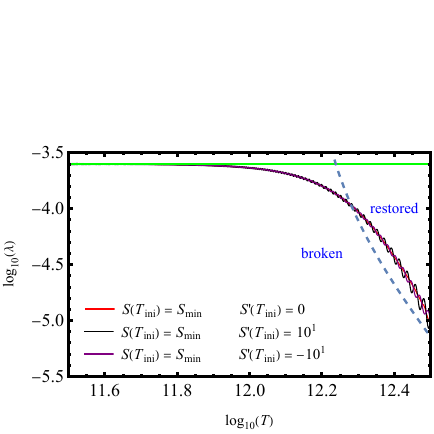}
    \caption{Sensitivity to initial conditions. The red, green and blue lines are the same as in \Cref{fig:PT-Moduli}. The black and purple lines represent the quartic coupling as a function of temperature for different initial conditions.}
    \label{fig:Sensitivity}
\end{figure}

A possible concern would be that the variation in the strength of the phase transition observed in \Cref{fig:PT-Moduli} results from a fine-tuned dependence on the initial conditions specified in \Cref{eq:InitialConditions}.  In \Cref{fig:Sensitivity}, we show that this is not the case by plotting the solution to the equation of motion in \eqref{eq:PhiMotionT} for initial conditions perturbed from those in \Cref{eq:InitialConditions}. In the left panel, we change the initial conditions by perturbing the solution around the instantaneous minimum of the effective potential at the initial temperature $T_\mathrm{ini}$. In the right panel, we change the initial conditions in the derivative of $\WKS$ at $T = T_\mathrm{ini}$. 
The perturbed initial conditions lead to solutions exhibiting significantly more pronounced oscillations. Nevertheless, the overall picture remains unchanged: the weakly coupled scalars $\WKS$ feel a force from the thermal bath, yielding a smaller quartic coupling at high temperatures. Hence, the first-order phase transition tends to become stronger.

\subsection{Thermal Effects in More Sophisticated Scenarios}
\label{subsec:More_sophisticated_scenarios}

We have argued that the thermal effects drive the weakly coupled scalar(s) $\WKS$ to values at which the interaction strengths decrease. 
There are exceptions to this statement. 
For instance, if one has more than one thermal bath, these effects may compete with each other. 
This happens when, say, increasing the thermal \ac{VEV} of $\WKS$ increases the interactions of one sector while decreasing the couplings in another sector. 
This situation is realized in the heterotic string, where the K\"ahler moduli enter the gauge kinetic function with opposite signs \cite{Ibanez:1986xy,Dixon:1990pc,Stieberger:1998yi}. 

The effects we are discussing here have some overlap with the paradigm of symmetry restoration at high temperature. 
It is well-known that there are exceptions to this paradigm, see e.g.~\cite{Jansen:1998rj}.
Under certain circumstances, there are more light degrees of freedom at finite couplings. 
In such scenarios, the free energy may favor larger couplings.  

As already discussed in \cite{Fardon:2003eh}, the sector which we have been referring to as ``thermal'' does not \emph{need} to be in thermal equilibrium. 
For instance, after \ac{BBN} neutrinos are no longer in equilibrium, yet they are expected to have an almost thermal spectrum corresponding to a temperature $T_\nu$.
Our discussion then applies in that scenario as well, and may also be generalized to non-thermal distributions. 

\subsection{Implications for String Cosmology}
\label{sec:string_cosmology}

In string theory couplings depend on weakly coupled fields, the so-called moduli.
These moduli correspond to classically flat directions and typically have a shallow potential. 
This leads to certain challenges for cosmology \cite{Coughlan:1983ci,deCarlos:1993wie,Kane:2015jia}. 
These so-called moduli problems include the following items:
\begin{enumerate}
 \item Moduli may get driven unphysical values such as the Dine--Seiberg runaway vacuum \cite{Dine:1985he}. This may be a consequence of moduli dynamics after inflation as in the Brustein--Steinhardt problem \cite{Brustein:1992nk} or due to the thermal effects discussed in this study \cite{Buchmuller:2004xr}, or due to the large energy density carried by the inflaton~\cite{Kallosh:2004yh}.
 \item Late decays of the moduli may be inconsistent with observations of \ac{BBN} \cite{Coughlan:1983ci,deCarlos:1993wie,Kawasaki:1995cy} or the \ac{CMB} \cite{Moroi:2001ct}. 
 \item Entropy produced in moduli decays before \ac{BBN} may dilute the \ac{BAU} and/or drive the \ac{DM} density to unrealistic values. 
\end{enumerate}
This last issue can be avoided if the moduli themselves are responsible for generating the \ac{BAU} and/or \ac{DM} relic abundance \cite{Kitano:2008tk,Chen:2018uzu,Kane:2019nes,Baer:2023bbn}.\par

Note that the thermal effects may not only lead to new challenges but may also play a role in finding new solutions to the moduli problems. 
For instance, in \cite{Dienes:2015bka} it has been proposed that moduli can be converted into each other resonantly. 
This requires their effective masses to cross. 
The thermal corrections studied in this work can lead to this situation if one modulus is more strongly coupled to the thermal bath than the other. 
In this case, one may hope to convert more weakly coupled, and thus more dangerous, moduli into more strongly coupled moduli which decay earlier and thus pose a reduced threat to cosmology.

\section{Conclusions}
\label{sec:Conclusions}

Weakly coupled scalars are generally out-of-equilibrium in the early Universe. 
However, they may nonetheless receive thermal corrections to their potential, which are different from the thermal potential of the equilibrated particles. 
This thermal dependence can drive the weakly coupled scalars to different expectation values, which in the case of a scalar acting as a modulus
can result in a change to the effective couplings of the theory. 
We have shown that for masses, gauge, and Yukawa couplings, the free energy exerts a force in the weakly coupled scalars which tends to drive 
the effective couplings to smaller values at high temperatures, and we highlight a novel application where the scalar quartic coupling becomes smaller at high temperatures. 
In this latter case, the symmetry-breaking phase transition can become stronger in the context of a Higgs particle that breaks a $\U{1}^{\prime}$ gauge symmetry. 
This shift in the quartic coupling could have observational effects, such as an impact on the gravitational wave spectrum produced by the phase transition.
Furthermore, apart from a potential modulus-dependent cubic term, this work provides a comprehensive picture of renormalizable couplings controlled by moduli. 
It would be interesting to investigate the detailed behavior of the cubic term, which we leave for future work.

\section*{Acknowledgements}

We thank Yang Bai and Jessica Turner for useful discussions. 
During the completion of this work, GM was supported in part by the UC office of the President through the UCI Chancellor's Advanced Postdoctoral fellowship, the U.S.\ National Science Foundation under Grant PHY-2210283, a UCI School of Physical Sciences Visiting Fellowship and the Natural Sciences and Engineering Research Council of Canada (NSERC). 
The work of VKP, MR and TT is supported in part by the U.S.\ National Science Foundation under Grant PHY-2210283.

\appendix
\section{Effective Potential in the Abelian Higgs Model}
\label{sec:EffectivePotentialHiggs}

In this Appendix, we compute the free energy of an equilibrated scalar that spontaneously breaks a $\mathrm{U}(1)^{\prime}$ gauge symmetry. We closely follow~\cite{Carrington:1991hz} but adapt their results to a $\mathrm{U}(1)^{\prime}$ gauge symmetry instead of the \ac{EW}. We start with the Lagrange density given by \Cref{eq:LbeforeSB}. Expanding the scalar field around the constant value $v$ 
\begin{equation}
    h = \frac{1}{\sqrt{2}}\left( h_1 + \ii h_2 + v \right) \; ,
    \label{eq:Expandh}
\end{equation}
and inserting \Cref{eq:Expandh} into \Cref{eq:LbeforeSB} gives
\begin{align}
    \mathscr{L} &= -\frac{1}{4}F^{\prime}_{\mu\nu}F^{\prime \mu\nu} + \frac{1}{2}\partial_\mu h_1\, \partial^\mu h_1+ \frac{1}{2}\partial_\mu h_2\, \partial^\mu h_2 \nonumber \\ 
    &\hphantom{{}={}}{}+\frac{1}{2}g^2\, v^2\, A^{\prime}_\mu A^{\prime \mu} - \frac{1}{2}\left(3\lambda\, v^2 -\mu^2\right)\,h_1^2 - \frac{1}{2}\left(\lambda\, v^2 -\mu^2\right)\,h_2^2\nonumber \\
    &\hphantom{{}={}}{}+ \frac{g^2}{2}A^{\prime}_\mu A^{\prime \mu}\,\left( h_1^2 + h_2^2 + 2 h_1\, v \right) - g\, A^{\prime \mu}\, h_1\,\partial_\mu h_2 + g\, A^{\prime \mu}\, h_2 \,\partial_\mu h_1 - g\, A^{\prime \mu}\, v\, \partial_\mu h_2 \nonumber\\
    &\hphantom{{}={}}{} - \lambda\, h_1\, v\, \left( h_1^2 +h_2^2\right) - \frac{\lambda}{4}\left(h_1^2+h_2^2\right)^2 - v\, \left( \lambda\, v^2 - \mu^2 \right)\, h_1 + \frac{\mu^2}{2}v^2 - \frac{\lambda}{4}v^4 + \text{gauge fixing terms}\; .
    \label{eq:LafterSB}
\end{align}
Thus, we see that the masses of the gauge boson $A^{\prime}_\mu$, the scalar field $h_1$ and $h_2$ are given by
\begin{subequations}    
\begin{align}
    m_{A^{\prime}}^{2}(v,g) &= v^2\, g^2\; ,  \\ 
    m_1^2(v, \lambda, \mu) &= 3\lambda\, v^2 - \mu^2 \; , \\
    m_2^2(v, \lambda, \mu) &= \lambda\, v^2 - \mu^2 \;  ,
\end{align}
\end{subequations}
respectively. 
The tree-level potential is given by
\begin{equation}
    \mathscr{V}_{\tree}^{(h)}(v,\mu, \lambda) = -\frac{\mu^2}{2}v^2 + \frac{\lambda}{4}v^4\; .
    \label{eq:Vtree}
\end{equation}

\begin{figure}
    \centering
    \subcaptionbox{1-loop contribution to the effective potential.\label{fig:Diagram1}}
    {\includegraphics[width=0.2\linewidth]{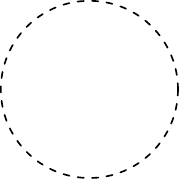}}\qquad
    \subcaptionbox{2-loop contribution to the effective potential due to quartic.\label{fig:Diagram3}}{\includegraphics[width=0.2\linewidth]{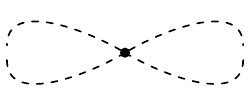}}
    \caption{Contributions to the effective potential.}
\end{figure}

The effective potential of an equilibrated scalar receives a $1$-loop contribution from the diagram in \Cref{fig:Diagram1}. It is given by~\cite[Equation (3)]{Carrington:1991hz}
\begin{align}
    \mathscr{V}_{1}(v,T) &  = \frac{T}{2}\sum_n \int\!\frac{\dd^3 k}{(2\pi )^3}\ln \left( \omega_n^2 + \vec{k}^2 + m^2(v) \right)\nonumber \\ 
    & \eqdef  \mathscr{V}_{1}^{(0)} (v) + \mathscr{V}_{1}^{(T)}(v,T) \; ,
\end{align}
where $\mathscr{V}_{1}^{(0)} (v)$ is the zero-temperature part and $\mathscr{V}_{1}^{(T)} (v)$ is the temperature-dependent part. 
The zero-temperature and temperature-dependent contributions to the effective potential are given in~\cite[Equations (4) and (7)]{Carrington:1991hz} and we rewrite them here as
\begin{subequations}\label{eq:1Loop}
\begin{align}
    \mathscr{V}_{1}^{(0)}(v) 
    &= \frac{1}{4\pi^2}\int_0^{\infty}\!\dd k\, k^2 \,\left( k^2 + m^2 (v) \right)^{1/2}\; ,\\
     \mathscr{V}_{1}^{(T)}(v,T) 
    &= \frac{T}{2\pi^2} \int\limits_{0}^\infty\!\dd k\, k^2\, \ln \left( 1 - \exp\left[ -\frac{1}{T} \left( k^2 + m^2 (v) \right)^{1/2}\right]\right) \eqdef J\bigl(m(v),T\bigr) \; .
\end{align}
\end{subequations}
Here we have defined the function $J(m,T)$ as
\begin{equation}
    J(m,T) = \frac{T^4}{2\pi^2}\, \int_0^\infty\!\dd x\,x^2\, \ln \left( 1- \exp\left[ -\sqrt{x^2+\left( \frac{m}{T}\right)^2}\right]\right)\;.
\end{equation}
Ignoring the zero-temperature part, the $1$-loop contributions to the effective potential are obtained from \Cref{eq:1Loop}. 
Furthermore, we include thermal corrections to the masses~\cite[Equations (28) and (37)]{Carrington:1991hz}, which are given by
\begin{subequations}
\begin{align}
    \Pi_{h_i} (T,g,\lambda) &= \frac{\lambda }{2}T^2 + \frac{g^2}{16}T^2 \;  ,  \\
     \Pi_{A^{\prime}} (T, g ) & =\frac{g^2}{6}T^2 \; ,
\end{align}
\end{subequations}
for the gauge boson $A^{\prime}$. 
Here $h_i$ for $i = 1,2$ label the two independent Higgs fields defined in \eqref{eq:Expandh}.
Furthermore, we include the leading term of the $2$-loop contribution from the two Higgs fields $h_i$ to the effective potential. 
This contribution arises from the diagram in \Cref{fig:Diagram3} and was computed in~\cite[Equation (3.78)]{Laine:2016hma}. 
To leading order, this is given by
\begin{equation}\label{eq:2-loop}
    \mathscr{V}_{2,h}^{(T)}(v, T,\lambda) = 2 \times\frac{3 \lambda}{4} \frac{T^4}{144}=\frac{\lambda}{96}T^4 \; ,
\end{equation}
where the factor of $2$ arises from the fact that we have two Higgs fields in \eqref{eq:Expandh}. 
Thus, the effective potential is given by
\begin{align}
 \MoveEqLeft   \mathscr{V}_{\eff}^{(h)} (v, T,\mu, \lambda, g)  
 = -\frac{\mu^2}{2}v^2 + \frac{\lambda}{4}v^4\nonumber\\ 
    &
    {}+J\bigl(m_{A^{\prime}}(v)+\Pi_{A^{\prime}} (T),T\bigr)+J\bigl(m_1(v)+\Pi_1 (T),T\bigr)+
    J\bigl(m_2(v)+\Pi_2 (T),T\bigr)
    +\frac{\lambda}{96}T^4\; .
    \label{eq:VeffAppendix}
\end{align}
We note that additional two-loop contributions to \Cref{eq:VeffAppendix} arise from the terms $A^{\prime}_\mu A^{\prime \mu} h_1^2$, $A^{\prime}_{\mu} A^{\prime \mu} h_2^2$, $A^{\prime}_\mu A^{\prime \mu} h_1$, and $h_1 \left( h_1^2 + h_2^2 \right)$. Among these, the contributions not proportional to $\lambda$ do not impact the equation of motion for $\WKS$ in \Cref{eq:EqMotionphi}. However, we acknowledge that the $\lambda$-dependent diagrams can modify the equation of motion. Despite this, the central message of this work remains unaffected: our goal is not to compute the precise numerical shift in the quartic coupling, but to demonstrate qualitatively that the presence of $\WKS$ modifies the strength of the phase transition.

\bibliography{KMRT}
\bibliographystyle{utphys}

\begin{acronym}
    \acro{BAU}{baryon asymmetry of the Universe}
    \acro{BBN}{big bang nucleosynthesis}
    \acro{CMB}{cosmic microwave background}
    \acro{DM}{dark matter}
    \acro{EW}{electroweak symmetry}
    \acro{SM}{standard model}
    \acro{VEV}{vacuum expectation value}
\end{acronym}

\end{document}